\documentclass[twocolumn,showpacs,preprintnumbers,amsmath,amssymb,showlabels,longbibliography]{revtex4-1}

\usepackage{graphicx}
\usepackage{dcolumn}
\usepackage{bm}

\begin{document}

\title{Symmetric wetting heterogeneity suppresses fluid displacement hysteresis in granular piles}

\author{R. Moosavi, M. Schr\"oter, S. Herminghaus}

\affiliation{Max-Planck-Institute for Dynamics and Self-Organization, Bunsenstr. 10, 37073 G\"ottingen, Germany}

\date{\today}

\begin{abstract}
We investigate experimentally the impact of heterogeneity on the capillary pressure hysteresis in fluid  invasion of model porous media. We focus on `symmetric' heterogeneity, where the contact angles the fluid interface makes with the oil-wet ($\theta_1$) and the water-wet ($\theta_2$) beads add up to $\pi$. While enhanced heterogeneity is usually known to increase hysteresis phenomena, we find that hysteresis is greatly reduced when heterogeneities in wettability are introduced. On the contrary, geometric heterogeneity (like bi-disperse particle size) does not lead to such effect. We provide a qualitative explanation of this surprising result, resting on rather general geometric arguments.

\end{abstract}


\maketitle

\section{Introduction}

Immiscible displacement of one fluid by another within a permeable medium, such as soil \cite{Wooding1997,Bocxlaer2011}, a porous rock  formation\citep{Anderson1987a,Morrow1990,Dullien1991,Sahimi2011}, or a filter cake \cite{Wong1999}, is of tremendous importance in environmental,  geological and industrial settings. A deeper understanding of the fundamental mechanisms governing the morphology of the propagating fluid front would be of immediate impact on, e.g.,   irrigation, oil recovery \citep{Ian2008,Austad2012,Hassenkam2011} and storage in  geological formations \citep{Anderson1987a, Morrow1990,Orr2009,Bickle2009}. However, progress in this direction has been strongly impeded by the enormous complexity inherent to these systems owing to their random geometry. This pertains not only to natural samples. Even in piles of perfectly spherical particles of equal size, which represent the most frequently studied model of a permeable solid, the random nature of the packing geometry renders the morphology of the invading fluid front poorly predictable. Nevertheless, it has been recently shown that important aspects of the morphology of the invading fluid front can be understood on the basis of merely the wetting geometry at the pore scale  \cite{Singh2017}.

In the present paper, we apply this paradigm to the impact of sample heterogeneity. Following established schemes \citep{Murison2014,Blunt1991,Blunt1997,Dixit2000,Oren2003,Zhao2010}, we study fluid displacement in samples consisting of piles of spherical glass beads with sub-millimeter diameter, which are  chemically functionalized in order to control their wettability. Furthermore, we study the interplay between geometric heterogeneity and heterogeneous wettability. Indications that correlations between  geometry and wettability  may play a role date back to 1959 \cite{Fatt1959} and should therefore be considered.

\section{Experiments}

Soda lime glass beads were purchased from MoSci Corporation and cleaned with piranha solution before further treatment. Oil-wet beads were created by liquid-phase silanization with octadecyl-trichlorosilane (OTS). The procedure consisted of  immersing the beads for $ 45 $ minutes in a solution prepared from $ 50\ \text{ml} $ bicylohexyl ($\textsf{C}_{12}\textsf{H}_{22}$), $ 0.240\ \text{ml}$ carbon tetrachloride $\textsf{C}\textsf{Cl}_4$ and $ 0.214\  \text{ml}$ of the OTS \citep{Lessel2014}. This was followed by a chloroform rinse to remove unbound OTS molecules. Subsequently, the sample was rinsed with ethanol and acetone and then dried in an oven for $30$ minutes. 
 
The procedure was repeated three times in order to enhance the quality of  the chemisorbed silane layer. On standard glass slides, this was found to yield an advancing contact angle for water against air of $ 114^{\circ}$, which underscores the high grafting density achieved \cite{Lessel2014}. 
 
\begin{figure}[h]
\centering
\includegraphics[width=\columnwidth]{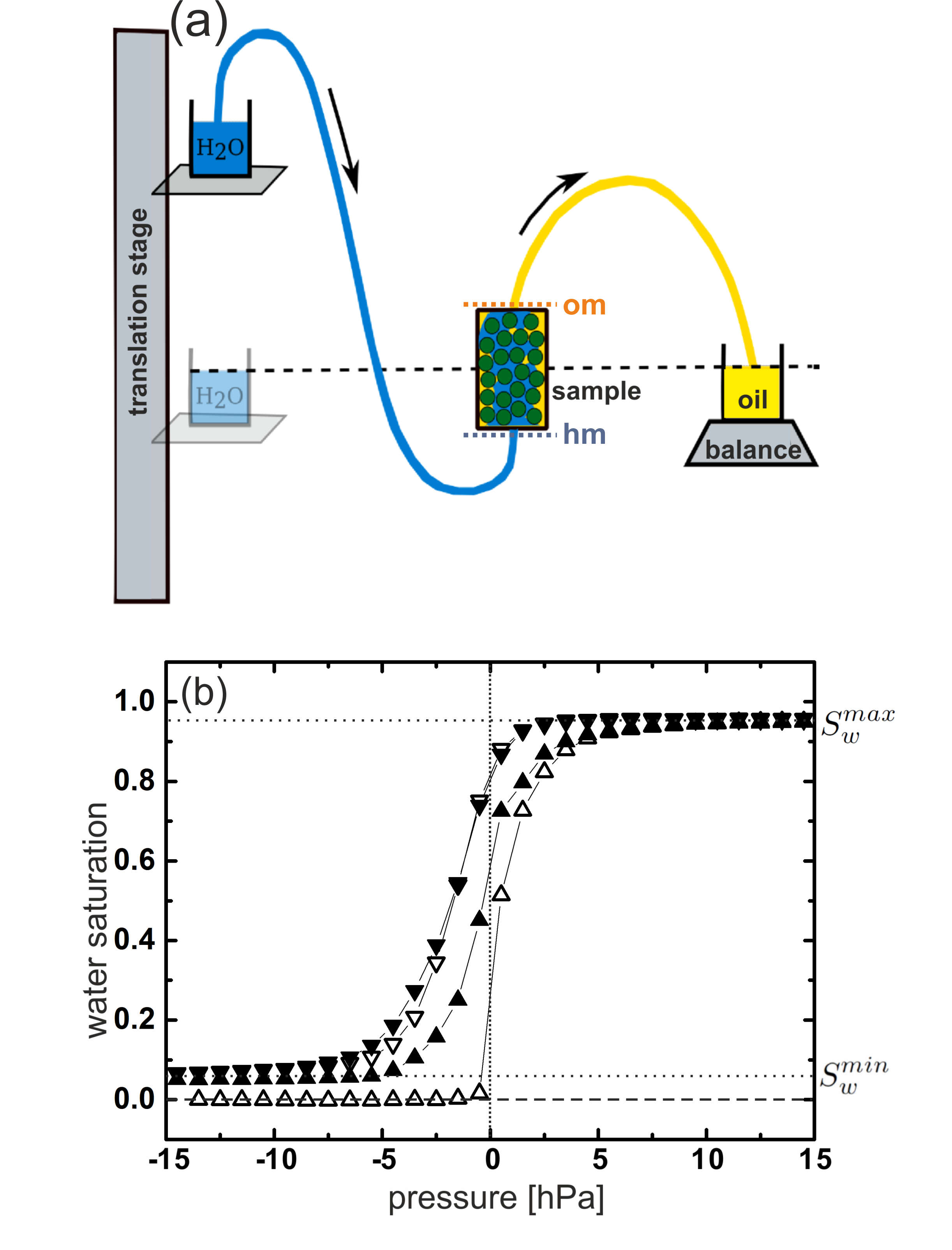}
\caption{{\bf (a)} Sketch of the experimental apparatus. The sample cell is sealed at the top and the bottom with an oleophilic ($\textsf{om}$) and a hydrophilic ($\textsf{hm}$) membrane, respectively. This prevents the fluid interface from leaving the cell. The hydrostatic pressure difference between water and oil is controlled by the height of the water reservoir using the translation stage. The mass of the displaced oil (and thereby the water saturation) is measured by a high precision balance. {\bf (b)} An example saturation curve. Open triangles: first run. Solid triangles: all consecutive runs.}
\label{Fig:ExpSetup}
\end{figure}

Our experimental setup is sketched in Fig.~\ref{Fig:ExpSetup}a. The bottom and top of the sample cell consist of a hydrophilic and oleophilic porous membrane, respectively, in order to prevent the fluid front from leaving the cell, and to thus keep the two liquids strictly separated outside the cell.  The hydrophobic (oleophilic) membrane at the top is made of $135\ \mu m$ thick porous Teflon, the one at the bottom is made of Polyamid (both supplied by Sartorius AG). As immiscible fluid phases, we chose hexadecane (Alfa Aesar, density $ \rho = 0.77\  \text{g}/\text{cm}^{-3} $), which was filtered prior to use through an  alumina column (height 20 cm) to remove surfactants, and milipore water as the aqueous phase. The interfacial tension between the two liquids was measured by means of the pendant drop method and found to be 52 $\pm 1$ mN/m. The particularity of this choice of liquids lies in the fact that the contact angles the fluid interface makes with the bead surfaces are lying symmetrically around $\pi/2$.  Experimental evidence and physical consequences will be  discussed further below.   
 
Samples are prepared by first pouring dry mixtures of beads into the oil phase, then thoroughly degassing. Subsequently the oil saturated with beads is filled into the sample cell. After sealing the cell, its top outlet is connected to a tube which ended in a beaker with the same fluid. The latter was positioned on a precision balance which allows to measure the amount of expelled oil. The bottom of the sample container is connected, with a flexible tube, to the bottom of a beaker containing water, which can be shifted vertically by means of a translation stage. This allows for precise adjustment of the pressure of the invading fluid (i.e., water). 
  
Capillary pressure saturation (CPS) curves were obtained with a home-built experimental setup which has been described in detail before \cite{Murison2014}. Each CPS measurement begins with the sample fully oil-saturated ($ \text{S}_{w}$ = 0 in Fig.~\ref{Fig:ExpSetup}b). The water is made to invade the cell by vertically translating the water reservoir in steps  equivalent to 1 mbar (upward pointing open triangles in Fig.~\ref{Fig:ExpSetup}b). At each incremental step, the water saturation $ \text{S}_{w}$ is calculated from the mass of displaced oil from the sample. After  oil saturation has been reached for the first time ( $S_w^{max}$), further increase of pressure does not produce more oil, because the fluid front cannot pass the oleophilic membrane at the top of the cell. The capillary pressure is then decreased again stepwise back to its original value, thereby allowing the oil phase to re-invade the sample cell (downward pointing open triangles in Fig.~\ref{Fig:ExpSetup}b). Subsequently, this procedure is cycled several times in order to establish a steady CPS curve (solid triangles  in Fig.~\ref{Fig:ExpSetup}b). This CPS curve was reproducible within experimental scattering. In a recent study \cite{Singh2017} with the same system, where the fluid interface was monitored by means of x-ray tomography, we could establish that gravity has no noticeable influence on the interface morphology and the behaviour of the system.

\begin{figure}[htbp]
\centering
\includegraphics[width=0.85\columnwidth]{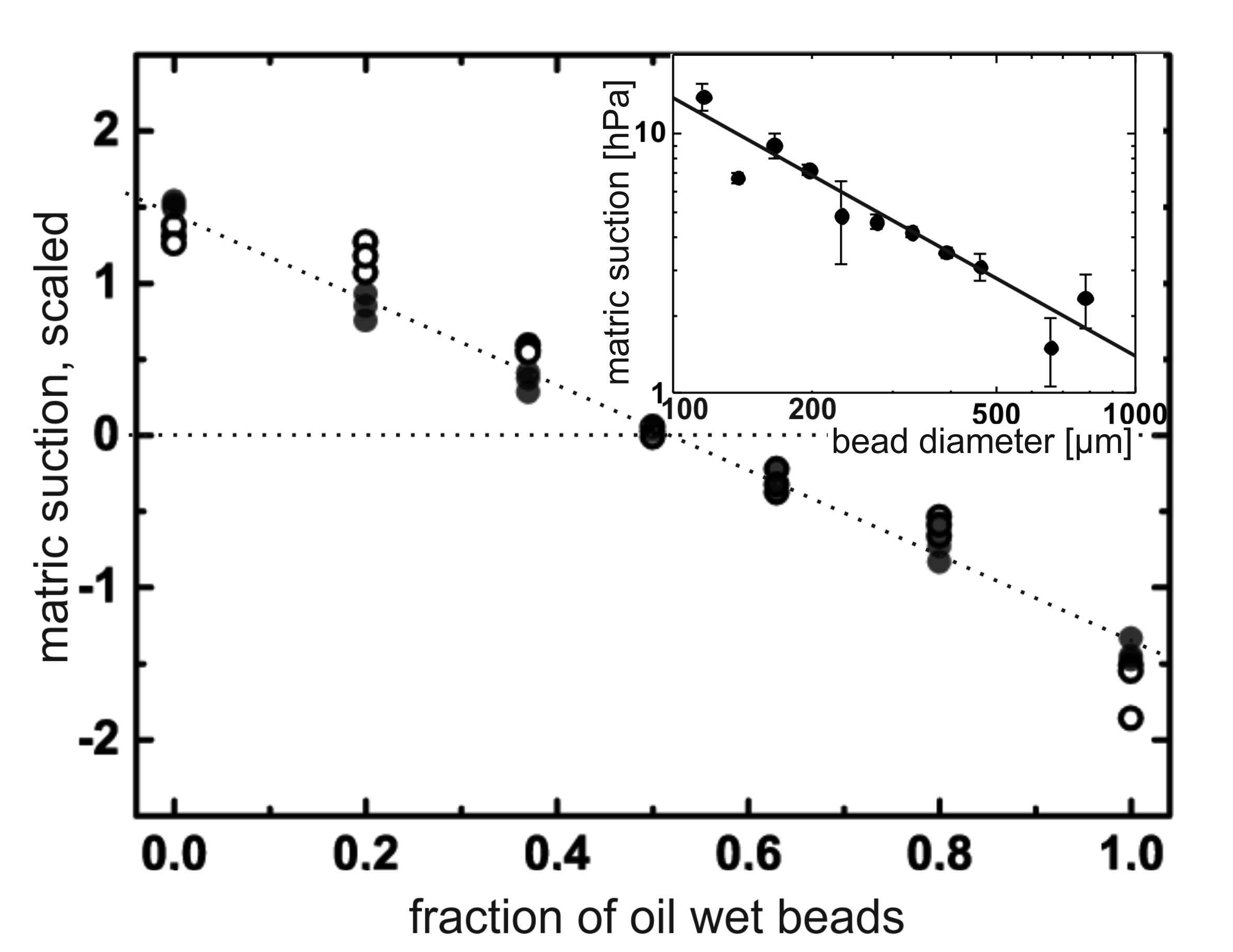}
\caption{Matric suction measured with samples of mixed wet beads of equal size. The scaled quantity $p_s R/\gamma$ is plotted on the ordinate, the abscissa is $\phi_{\textrm{oi}l}=\phi_2=(\Phi_2-\Phi_1)/2\Phi+\frac{1}{2}$. Closed circles: $R=116\pm 9 \mu$m. Open circles: $R=195\pm16 \mu$m. {\it Inset}: log-log plot of $p_s$ vs. the bead size, measured with mono-disperse samples of water-wet beads. The straight line has unity slope.}
\label{Fig:SuctionMonodisperseMixedWet}
\end{figure}
 
\section{Results}

When the interface  between two fluids is in contact with a solid surface, the equilibrium interface morphology is characterized by a certain contact angle which establishes itself between the fluid interface and the solid surface. If $\gamma_{1}$ and $\gamma_{2}$ are the interfacial tensions between fluid $1$ or fluid $2$ with the solid, respectively, the equilibrium contact angle is given by
\begin{equation}
\gamma \cos \theta = \gamma_1-\gamma_2
\label{Eq:Young}
\end{equation}
where $\gamma$ is the tension of the interface between the two (assumed immiscible) liquids. If  a fluid is being completely expelled by a second, immiscible  fluid, the corresponding capillary pressure, $p_s$, is given by the excess interfacial free energy gain upon exchanging one fluid with the other. It is directly proportional to the difference in the interfacial tensions, $\gamma_i$, and hence to $\cos\theta$, and is commonly called the {\it matric suction}.  
 
For a sample of equally sized spherical beads with contact angle $\theta$ and radius $R$,   one readily obtains
\begin{equation}
p_s = \frac{\Phi}{1-\Phi}\frac{3\gamma \cos\theta}{R}, 
\label{Eq:SuctionSimple}
\end{equation}
where $\Phi$ is the filling fraction of the beads (fraction of total sample volume) and $R$ is their radius. That the matric suction indeed scales with the inverse radius of the beads in our experiments is shown in the inset of Fig.~\ref{Fig:SuctionMonodisperseMixedWet}.

For a sample containing two types of beads with  different wettability  (i.e., different contact angles $\theta_1$ and $\theta_2$) and different size, we obtain
\begin{equation}
p_s = \frac{3\gamma}{1-\Phi}\left( \frac{\Phi_1 \cos\theta_1}{R_1} + \frac{\Phi_2 \cos\theta_2}{R_2}\right),
\label{Eq:SuctionBidisperseMixed}
\end{equation}
where $\Phi_1+\Phi_2=\Phi$.
This suggests that in a sample containing a mixture of water-wet and oil-wet beads, we should find a linear variation of $p_s$ with the relative content of the bead surface area. Accordingly, we plotted in the main panel of Fig.~\ref{Fig:SuctionMonodisperseMixedWet} the matric suction as a function of the fraction of the oil wet beads in mixed wet samples. The data of the filled and open symbols have been obtained with beads of different sizes and scaled with respect to $\gamma/R$, according to eq.~(\ref{Eq:SuctionSimple}). Not only do we find a reasonable data collapse, but also we indeed find a linear relationship  within experimental scattering. That $p_s$ goes to zero at equal fractions of both bead types indicates that $\cos\theta_1\approx -\cos\theta_2$. This is important to note because it suggests that the geometry of the invading and receding fronts should be identical. This is what we refer to as {\it symmetric wetting heterogeneity}. We then can write $\theta_{1/2}=\frac{\pi}{2}\pm\eta$, or $\theta_1+\theta_2=\pi$, and hence 
\begin{equation}
p_s = \frac{3\gamma\sin\eta}{1-\Phi}\left( \frac{\Phi_2}{R_2} - \frac{\Phi_1}{R_1}\right).
\label{Eq:SuctionBidisperseMixedOurcase}
\end{equation}
From the slope of the straight line in the main panel of Fig.~\ref{Fig:SuctionMonodisperseMixedWet} and from our system parameters ($\gamma = 52$mN/m, $\Phi = 0.60$) we obtain $\sin\eta=0.31$ and hence $\eta = 18 \ \textrm{deg}$.

Let us now turn to the hysteresis in the drainage curves, $\Delta p = p_+-p_-$, where  $p_+$ and $p_-$ are the critical pressures for invasion and for drainage, respectively. They correspond to the positions of the steep slopes of the saturation curves as displayed in Fig.~\ref{Fig:ExpSetup}. We define ($j\in\{+,-\}$)
\begin{equation}
p_{j} =  p_{max}\Delta S - \int\limits_{p_{min}}^{p_{max}}\left[S_w^{j}(p)-S_w^{min}\right]dp,
\label{Eq:CritPressures}
\end{equation}
where $p_{min}$  and  $p_{max}$ are chosen sufficiently far outside the range displayed in Fig.~\ref{Fig:ExpSetup}b to assure convergence of the branches of the saturation curves. $S_w^+(p)$ and $S_w^-(p)$ are the water saturations for ascending (invasion) and descending (drainage) pressure, and $\Delta S =  S_w^{max} - S_w^{min}$ is the amount of reversibly expelled fluid. 

If a fluid interface moves with respect to a solid surface it makes contact with, one frequently observes a certain hysteresis in the contact angle, i.e., a difference between the contact angle measured with advancing front ($\theta_{adv}$) and the contact angle when measured with the receding front ($\theta_{rec}$). The reason for this hysteresis is the  heterogeneity of  the solid surface, be it topographical or chemical \cite{Perrin2016}. If heterogeneity increases, so does the contact angle hysteresis (CAH), $\theta_{adv}-\theta_{rec}$. \cite{Dufour2016,Wylock2012}. Intuitively, one expects a similar phenomenology for wetting of   inhomogeneous porous media. If we add heterogeneity to the porous medium, we might expect the hysteresis, $\Delta p$ (cf.~Fig.~\ref{Fig:ExpSetup}b), to increase. 
 
Quite surprisingly, however, we observe the opposite. While heterogeneity in geometry, as introduced by mixing beads of different size, is found to have no discernible impact on imbibition hysteresis, heterogeneity in wettability even dramatically {\it reduces} the hysteresis of fluid invasion. This is rather counterintuitive, and we will provide here not only the evidence but also a tentative explanation based on simple geometric arguments.

\section{Discussion}

Let us first discuss $\Delta p$ for binary mixtures of beads of different size but equal wettability (oil wet or water wet). There are two main contributions to this hysteresis. One is the amplitude of variations in the interfacial free energy as the fluid front advances within the pile, and has to accommodate varying geometries of the interstitial space. We will call this contribution the {\it geometry induced hysteresis}.  These pressure variations should scale strictly with the variations of the mean curvature of the fluid front, and hence  with the size of throats between adjacent beads. The throat size between beads of different radii $R_i$ can neither be expressed in terms of $\langle R_i \rangle$  nor in terms of $\langle 1/R_i \rangle^{-1}$. Nevertheless, $\langle R_i \rangle$  certainly provides a reasonable measure for the intrinsic length scale of the sample. Hence we will, in the present paper, generally use $\langle R \rangle$ as the characteristic length scale of the system. 
 
The other contribution is the hysteresis inherent to the (heterogeneous) solid surfaces of the beads themselves. As mentioned above, this gives rise to CAH, which  is expected to contribute additively to the hysteresis observed in the drainage curves. 
 
\begin{figure}[htbp]
\centering
\includegraphics[width=\columnwidth]{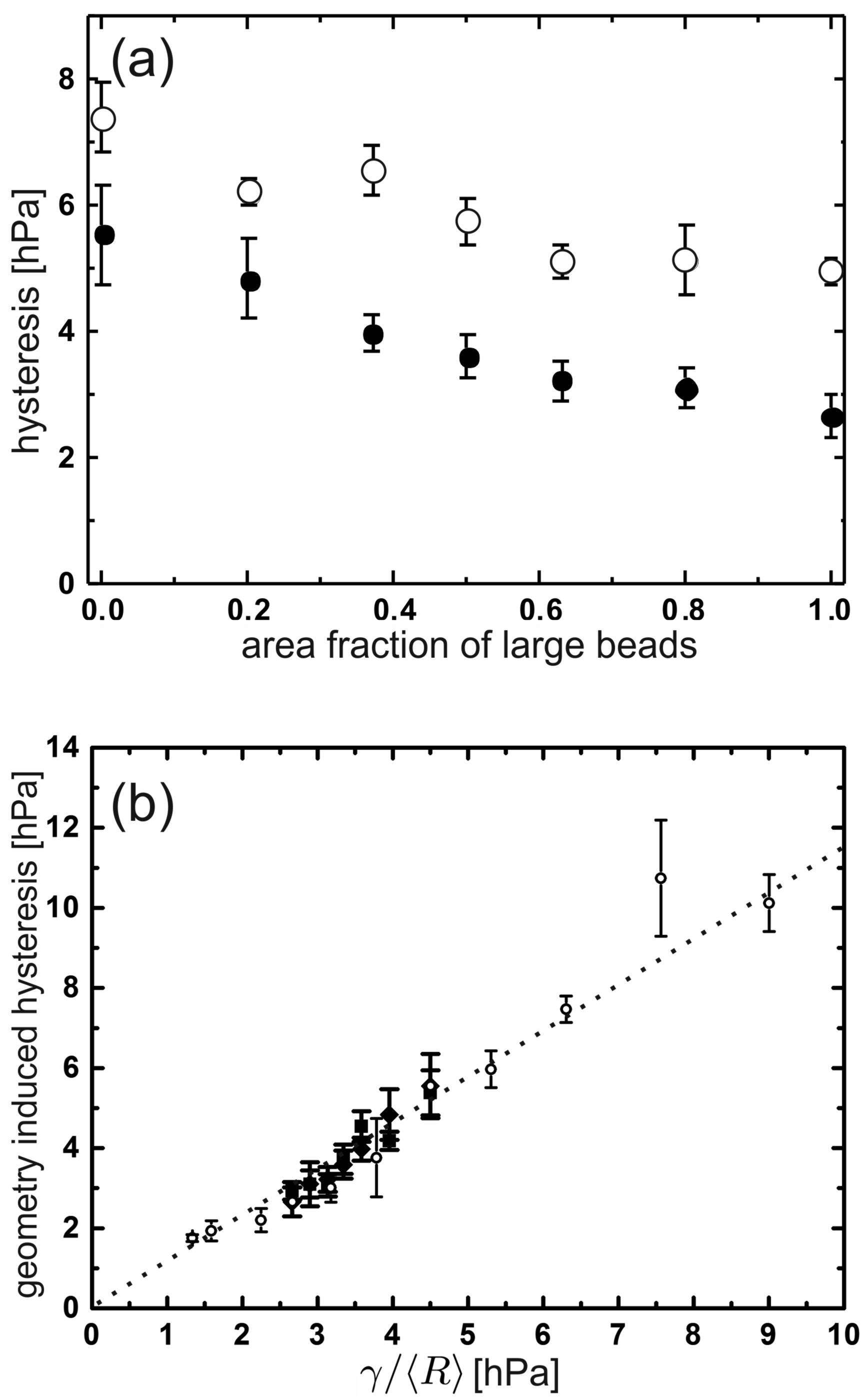}
\caption{{\bf (a)} Hysteresis measured with bi-disperse piles of oil wet (open symbols) and water wet (full symbols) beads, as a function of the relative fraction $\phi$ of large beads. Within scattering, the data sets are almost in parallel, the decreasing trend stemming from the change in average length scale (small beads and high pressures, left hand side, to large beads and low pressures, right hand side).   {\bf (b)} Hysteresis as a function of the capillary pressure scale, $\gamma/\langle R \rangle$. Open symbols:  monodisperse beads of variable size. Full symbols: bidisperse mixtures of beads. Circles: $R_1 = 116\mu$m, $R_2 = 195\mu$m. Triangles: $R_1 = 58\mu$m, $R_2 = 231\mu$m. The dotted line with unity slope represents what would be expected if the bead size were uniformly varying between the largest (left) and the smallest (right) size.}
\label{Fig:HystBidisperseAll}
\end{figure}

In fact, we observe that the hysteresis scales in a characteristic manner with the bead diameter for both water-wet and oil-wet beads, as shown in Fig.~\ref{Fig:HystBidisperseAll}a, where larger beads lead to smaller pressure hysteresis and vice versa. However, we also  observe a substantial additive difference of $2.03$ hPa on average between both data sets. If this is subtracted from the data obtained for oil wet samples, one finds a reasonable data collapse with hysteresis data obtained for monodisperse water wet samples of variable bead size, as shown in Fig.~\ref{Fig:HystBidisperseAll}b. The unity slope of the dotted line indicates that all pressures scale inversely proportional to the average size of the beads, which strongly suggests that this is entirely due to geometric effects, and not noticeably affected by CAH at the three-phase contact lines. 
 
We conclude that we have appreciable CAH on the surfaces of the oil wet beads, which gives rise to the extra $2.03$ hPa of hysteresis apparent in Fig.~\ref{Fig:HystBidisperseAll}a. This is well conceivable, as the oil wet beads had to be chemically functionalized, which tends to add some chemical heterogeneity, and hence CAH. This contribution will henceforth be called the {\it intrinsic hysteresis}. Although it is somewhat surprising that the CAH on the water-wet beads is too small to show up in our data, we have to accept this as an experimental result and will henceforth treat CAH on the water-wet beads as negligible for the effects to be considered here. 
 
The main conclusion we draw from Fig.~\ref{Fig:HystBidisperseAll}  is that geometric heterogeneity, which is maximal at equal fractions of large and small beads (center of Fig.~\ref{Fig:HystBidisperseAll}a), has no discernible effect on the hysteresis. Instead, the latter linearly interpolates between the extreme values corresponding to the mono-disperse samples, as supported by the data collapse in Fig.~\ref{Fig:HystBidisperseAll}b. 
   
\begin{figure}[htbp]
\centering
\includegraphics[width=\columnwidth]{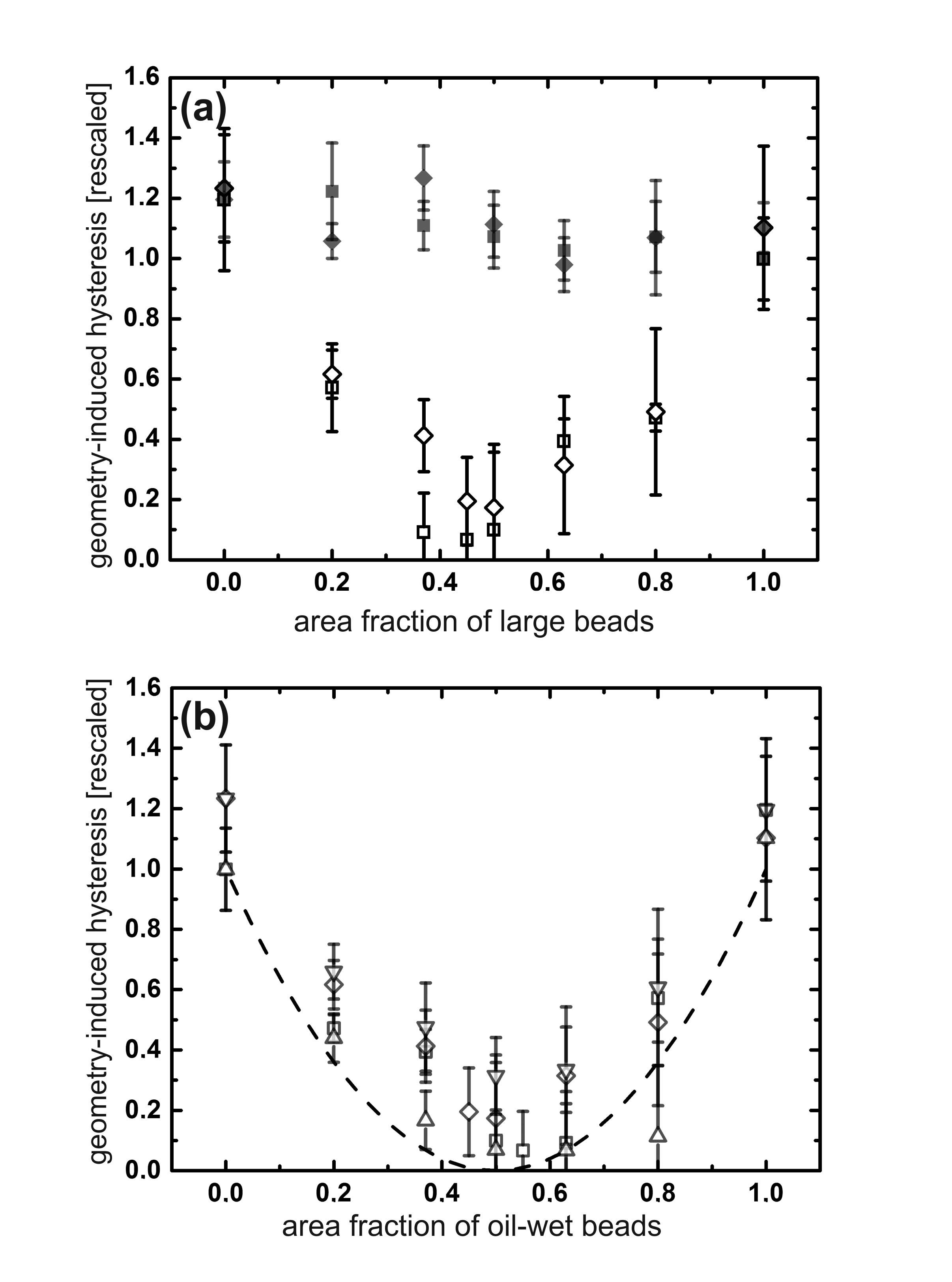}
\caption{Hysteresis measured with bi-disperse piles of beads, as a function of the relative fraction $\phi$. {\bf (a)} Mixtures of beads of different size ($R_1=116\pm9\ \mu$m, $R_2=195\pm16\ \mu$m). {\it Closed symbols:} beads of equal wettability. {\it Open symbols:} beads of different wettability. Squares: large beads water wet, small beads oil wet. Diamonds: large beads oil wet, small beads water wet. {\bf (b)} Mixtures of beads of different wettability. Squares and diamonds, data as above. Upward pointing triangles, large beads only. Downward pointing triangles, small beads only. Dashed curve, see text.}
\label{Fig:HystBidisperseMixedWet}
\end{figure}

Let us now include heterogeneity of wettability.  Fig.~\ref{Fig:HystBidisperseMixedWet} contains all data of geometry induced hysteresis for variable situations studied. For all data points involving oil-wet beads, the intrinsic hysteresis has been subtracted according to their fraction of sample surface.  Furthermore, data have been scaled the same way as in Fig.~\ref{Fig:SuctionMonodisperseMixedWet}, with $R$ having been replaced by the average radius, $\langle R\rangle$, as suggested by Fig.~\ref{Fig:HystBidisperseAll}b. The main observation is that while  find no noticeable influence of geometric heterogeneity on the width of the hysteresis, we observe an almost complete {\it suppression} of the hysteresis in the case of mixed wet piles, irrespective of the geometric heterogeneity.

\begin{figure}[htbp]
\centering
\includegraphics[width=\columnwidth]{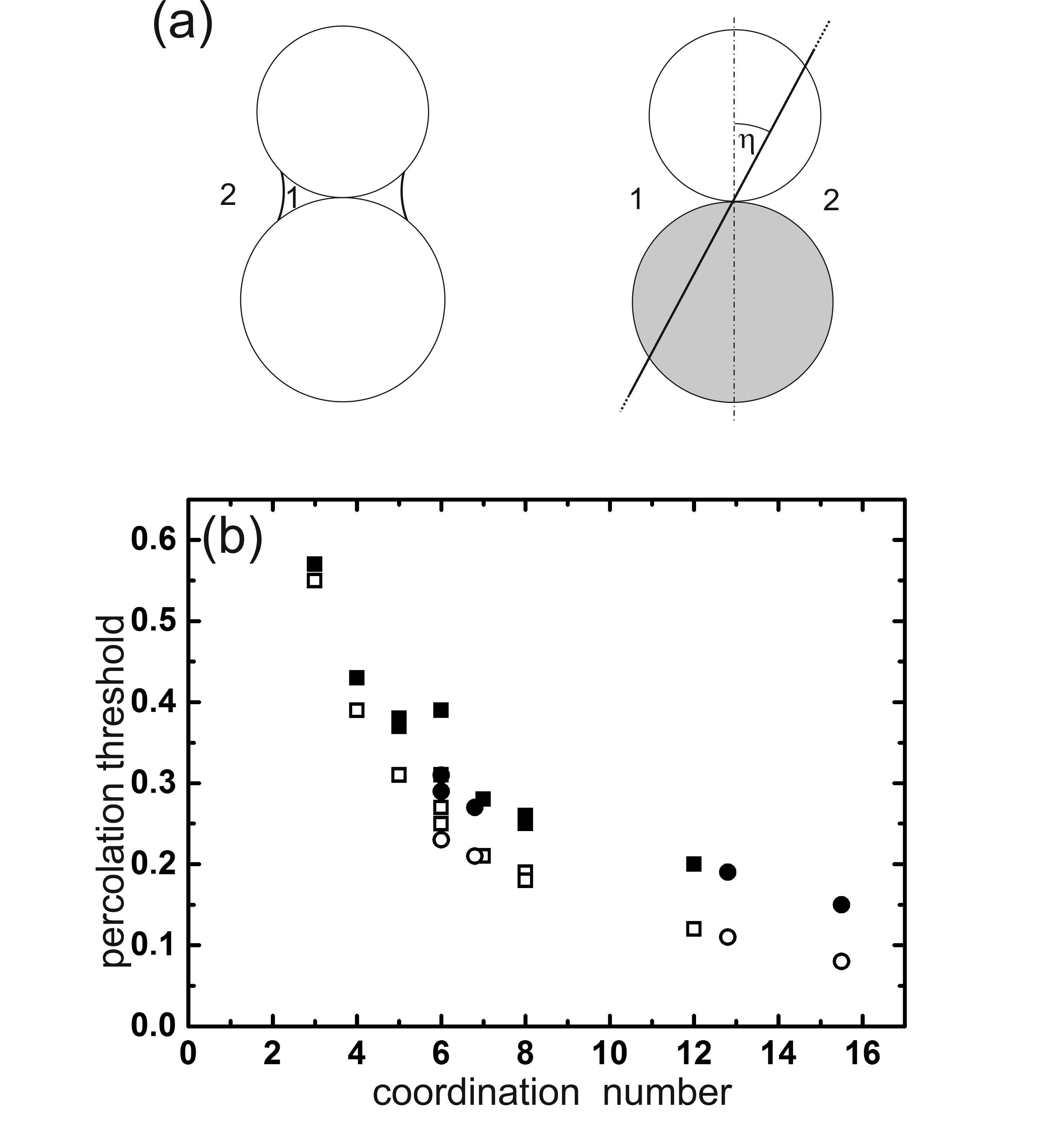}
\caption{{\bf (a)} Interface (solid) between two fluids (1 and 2) near a contact between two beads. {\it Left:} Contact between two beads wettable by fluid 1. The latter is likely to form a capillary bridge between both beads, with fluid 2 as the surrounding phase. {\it Right:} Contact between two beads of different wettability. A capillary bridge is not likely to form. For the `symmetric' case  $\theta_i=\pi/2\pm\eta$ (as for our system), there is not even an axisymmetric solution. Instead, the interface goes flat through the contact point, making an angle $\eta$ with the line connecting the centers of the beads (dash-dotted). {\bf (b)} The thresholds for site percolation (filled symbols) and bond percolation (open symbols) for a range of ordered (squares) and random (circles) lattices (\cite{Powell1979} and references therein).}
\label{Fig:Percolation}
\end{figure}

To arrive at a tentative explanation of this counter-intuitive result, we consider the geometry of the fluid interface close to a contact between two adjacent beads. If these have the same wettability, as sketched on the left hand side of Fig.~\ref{Fig:Percolation}a, the fluid which wets the beads better (fluid 1 in this case) will form a capillary bridge \cite{Kohonen2004}, with the other fluid as the surrounding phase. If the beads have different wettability, a capillary bridge of this kind is not likely to form. In particular, if the two contact angles are symmetric around $\pi/2$ as in our case ($\theta_i=\frac{\pi}{2}\pm\eta$ with $\eta > 0$), an axisymmetric solution like the one shown in the left sketch does not exist. Hence we have to discuss which fluid interface morphology is likely to emerge at a contact point between two spherical beads with contact angles $\theta_1$ and $\theta_2$, where $\theta_1+\theta_2\approx \pi$. 
 
It is straightforward to appreciate that the arrangement sketched on the right side of Fig.~\ref{Fig:Percolation}a provides a solution. It consists of a plane fluid interface which contains the point of contact and makes an angle $\eta = |\theta_i-\pi|$ with the line connecting the centers of the two beads (dash-dotted). One readily sees that this plane makes a contact angle of $\frac{\pi}{2}+\eta$ at its line of intersection with one of the beads,  and a contact angle of $\frac{\pi}{2}-\eta$ on its line of intersection with the other. As the flat fluid interface corresponds to zero Laplace pressure, this solution is expected to form easily, and not to add to the Laplace pressure constituting the geometric hysteresis of the CPS curve. It is expected to be pinned to the contact point by a mild minimum around zero Laplace pressure. Furthermore, it will provide a natural boundary between regions in which the liquid phase wetting the majority of beads represents the majority phase liquid. Excursions from zero applied pressure (cf.~Fig.~\ref{Fig:ExpSetup}b), in both positive and negative direction, should then easily displace fluid on the low pressure side.

On the other hand, the wetting phase will form capillary bridges at the contact points between like wet beads, as sketched in the left part of Fig.~\ref{Fig:Percolation}. From studies of wet granular piles \cite{Scheel2008,Scheel2008a,Herminghaus2013}, it is well known that neighboring capillary bridges start to grow into clusters already at a liquid content of about 2,5$\%$. For piles of beads of equal wettability, there is a bicontinuous phase in a wide range of liquid content, where both the wetting and the non-wetting phase are percolated \cite{Herminghaus2013}. The former as merged capillary bridges, the latter filling the interstitial space in between.  If  both fluids form percolated networks, they can in principle be depleted or inflated without any substantial movement of contact lines or any other rearrangements of fluid interface geometry. However, for piles with a single wettability, this is only possible in the late stage of expelling the wetting (i.e., capillary bridge) fluid \cite{Singh2017}. For most of the range of saturation, expelling one liquid with the other is accompanied by frequent rearrangements and concomitant changes in the topology of the fluid interface \cite{Scheel2008,Scheel2008a}, hence with considerable hysteresis in the CPS curve. 

In the mixed wet pile, however, the situation is different since both fluids have the same (or very similar) topology. It can be qualitatively discussed by considering the statistics of capillary bridges (or equivalent structures) occurring in the sample.  In a random pile of spheres, it is well known that there are about six contacts on each sphere \cite{Powell1979,Kohonen2004,Herminghaus2013}. Hence each contact has four nearest neighbor contacts on the same sphere. Consequently, each contact, and hence each capillary bridge, has $z = 8$ nearest neighbors in total on the two spheres between which it is formed.
 
In order to discuss our results in the context of percolation theory, we show in Fig.~\ref{Fig:Percolation}b the known percolation thresholds as a function of coordination number $z$ for several different lattices, including ordered and disordered ones. We see that aside from the data grouping into the two subgroups corresponding to bond percolation (open symbols) and site percolation (filled symbols), the percolation threshold is quite independent of the precise structure of the lattices, and can be well related to the coordination number as the only relevant descriptor. We realize that at $z = 8$ about 25$\%$ of occupancy is sufficient for site percolation. Hence for each of the liquid phases a percolated cluster of capillary bridges will persist as long as not more than 75$\%$ of the spheres are wet for the other liquid. This predicts that the effect of hysteresis suppression should be present around equal fractions of beads of both wettabilities, and extend from about 25$\%$ to  about 75$\%$. This is roughly what we observe in the experiments.
 
A complementary approach which leads to a similar result  may be gained from the number of oppositely wet contacts, as sketched on the right hand side of  Fig.~\ref{Fig:Percolation}a. If $\varphi_i$ is the number fraction of beads wet by fluid $i$, we have $\varphi_i+\varphi_2 = 1$, and the fraction of like-wet contacts is $\varphi_{l}= \varphi_1^2+\varphi_2^2=2\varphi_1^2-2\varphi_1+1$. The fraction of oppositely-wet contacts is $\varphi_{o}= 1-\varphi_l$. If we plot $\varphi_l-\varphi_o =  4\varphi_1^2-4\varphi_1+1$ into Fig.~\ref{Fig:HystBidisperseMixedWet} as a proxy for the necessity of  topologocal rearrangements (dashed curve), we see that the tendency observed in the CPS curve hysteresis is quite well reproduced.

\section{Conclusions and outlook}

Clearly, the details of the complex topology of the fluid interface still leaves a lot of room for investigations, but qualitatively the observed phenomenon seems to be captured by the above considerations.  One may think of a wide range of potential applications here. For instance, it seems possible to provide a novel design for  oil-water separation columns, where the interstitial geometry of the fine granulate provides a particularly large area of interaction between both phases. Furthermore, it will be interesting to investigate how universal the behaviour reported here turns out to be, in view of the rather simple mechanism behind it. In particular, it will be of great interest to investigate until how far away from symmetric heterogeneity this effect persists. In other words: how large has $\theta_1+\theta_2-\pi$ to be for this effect to break down?  Moreover, many hysteresis effects in fluid displacement in porous media  should be revisited in view of the present results.  
 
\section*{Acknowledgements}
 
The authors appreciate generous financial support from BP plc. within the ExploRe program.


\begin{thebibliography}{29}%
\makeatletter
\providecommand \@ifxundefined [1]{%
 \@ifx{#1\undefined}
}%
\providecommand \@ifnum [1]{%
 \ifnum #1\expandafter \@firstoftwo
 \else \expandafter \@secondoftwo
 \fi
}%
\providecommand \@ifx [1]{%
 \ifx #1\expandafter \@firstoftwo
 \else \expandafter \@secondoftwo
 \fi
}%
\providecommand \natexlab [1]{#1}%
\providecommand \enquote  [1]{``#1''}%
\providecommand \bibnamefont  [1]{#1}%
\providecommand \bibfnamefont [1]{#1}%
\providecommand \citenamefont [1]{#1}%
\providecommand \href@noop [0]{\@secondoftwo}%
\providecommand \href [0]{\begingroup \@sanitize@url \@href}%
\providecommand \@href[1]{\@@startlink{#1}\@@href}%
\providecommand \@@href[1]{\endgroup#1\@@endlink}%
\providecommand \@sanitize@url [0]{\catcode `\\12\catcode `\$12\catcode
  `\&12\catcode `\#12\catcode `\^12\catcode `\_12\catcode `\%12\relax}%
\providecommand \@@startlink[1]{}%
\providecommand \@@endlink[0]{}%
\providecommand \url  [0]{\begingroup\@sanitize@url \@url }%
\providecommand \@url [1]{\endgroup\@href {#1}{\urlprefix }}%
\providecommand \urlprefix  [0]{URL }%
\providecommand \Eprint [0]{\href }%
\providecommand \doibase [0]{http://dx.doi.org/}%
\providecommand \selectlanguage [0]{\@gobble}%
\providecommand \bibinfo  [0]{\@secondoftwo}%
\providecommand \bibfield  [0]{\@secondoftwo}%
\providecommand \translation [1]{[#1]}%
\providecommand \BibitemOpen [0]{}%
\providecommand \bibitemStop [0]{}%
\providecommand \bibitemNoStop [0]{.\EOS\space}%
\providecommand \EOS [0]{\spacefactor3000\relax}%
\providecommand \BibitemShut  [1]{\csname bibitem#1\endcsname}%
\let\auto@bib@innerbib\@empty
\bibitem [{\citenamefont {Wooding}\ \emph {et~al.}(1997)\citenamefont
  {Wooding}, \citenamefont {Tyler},\ and\ \citenamefont
  {"White}}]{Wooding1997}%
  \BibitemOpen
  \bibfield  {author} {\bibinfo {author} {\bibfnamefont {R.~A.}\ \bibnamefont
  {Wooding}}, \bibinfo {author} {\bibfnamefont {Scott~W.}\ \bibnamefont
  {Tyler}}, \ and\ \bibinfo {author} {\bibfnamefont {Ian}\ \bibnamefont
  {"White}},\ }\bibfield  {title} {\enquote {\bibinfo {title} {Convection in
  groundwater below an evaporating salt lake},}\ }\href@noop {} {\bibfield
  {journal} {\bibinfo  {journal} {Water Resources Research}\ }\textbf {\bibinfo
  {volume} {33}},\ \bibinfo {pages} {1199--1217} (\bibinfo {year}
  {1997})}\BibitemShut {NoStop}%
\bibitem [{\citenamefont {Bocxlaer}\ \emph {et~al.}(2011)\citenamefont
  {Bocxlaer}, \citenamefont {Verschuren}, \citenamefont {Schettler},\ and\
  \citenamefont {Kr\"opelin}}]{Bocxlaer2011}%
  \BibitemOpen
  \bibfield  {author} {\bibinfo {author} {\bibfnamefont {Bert~van}\
  \bibnamefont {Bocxlaer}}, \bibinfo {author} {\bibfnamefont {Dirk}\
  \bibnamefont {Verschuren}}, \bibinfo {author} {\bibfnamefont {Georg}\
  \bibnamefont {Schettler}}, \ and\ \bibinfo {author} {\bibfnamefont {Stefan}\
  \bibnamefont {Kr\"opelin}},\ }\bibfield  {title} {\enquote {\bibinfo {title}
  {Modern and eary {Holocene} mollusc fauna of the {Ounianga} lakes:
  implications for the paleohydrology of the central {Sahara}},}\ }\href@noop
  {} {\bibfield  {journal} {\bibinfo  {journal} {J. Quaternary Sci.}\ }\textbf
  {\bibinfo {volume} {26}},\ \bibinfo {pages} {433--447} (\bibinfo {year}
  {2011})}\BibitemShut {NoStop}%
\bibitem [{\citenamefont {Anderson}(1987)}]{Anderson1987a}%
  \BibitemOpen
  \bibfield  {author} {\bibinfo {author} {\bibfnamefont {W.~G.}\ \bibnamefont
  {Anderson}},\ }\bibfield  {title} {\enquote {\bibinfo {title} {Wettability
  literature survey- part 4: Effects of wettability on capillary pressure},}\
  }\href@noop {} {\bibfield  {journal} {\bibinfo  {journal} {Journal of
  Petroleum Technology}\ }\textbf {\bibinfo {volume} {39}},\ \bibinfo {pages}
  {1283--1300} (\bibinfo {year} {1987})}\BibitemShut {NoStop}%
\bibitem [{\citenamefont {Morrow}(1990)}]{Morrow1990}%
  \BibitemOpen
  \bibfield  {author} {\bibinfo {author} {\bibfnamefont {Norman~R.}\
  \bibnamefont {Morrow}},\ }\bibfield  {title} {\enquote {\bibinfo {title}
  {Wettability and its effect on oil recovery},}\ }\href@noop {} {\bibfield
  {journal} {\bibinfo  {journal} {Journal of Petroleum Technology}\ }\textbf
  {\bibinfo {volume} {42}},\ \bibinfo {pages} {1477--1484} (\bibinfo {year}
  {1990})}\BibitemShut {NoStop}%
\bibitem [{\citenamefont {Dullien}(1991)}]{Dullien1991}%
  \BibitemOpen
  \bibfield  {author} {\bibinfo {author} {\bibfnamefont {F.~A.~L.}\
  \bibnamefont {Dullien}},\ }\href@noop {} {\emph {\bibinfo {title} {Porous
  Media: Fluid Transport and Pore Structure}}}\ (\bibinfo  {publisher}
  {Academic Press},\ \bibinfo {year} {1991})\BibitemShut {NoStop}%
\bibitem [{\citenamefont {Sahimi}(2011)}]{Sahimi2011}%
  \BibitemOpen
  \bibfield  {author} {\bibinfo {author} {\bibfnamefont {M.}~\bibnamefont
  {Sahimi}},\ }\href@noop {} {\emph {\bibinfo {title} {Flow and Transport in
  Porous Media and Fractured Rock: From Classical Methods to Modern
  Approaches}}}\ (\bibinfo  {publisher} {John Wiley and Sons},\ \bibinfo {year}
  {2011})\BibitemShut {NoStop}%
\bibitem [{\citenamefont {Wong}(2013)}]{Wong1999}%
  \BibitemOpen
  \bibfield  {author} {\bibinfo {author} {\bibfnamefont {Po-Zen}\ \bibnamefont
  {Wong}},\ }\href@noop {} {\emph {\bibinfo {title} {Methods in the Physics of
  Porous Media}}},\ \bibinfo {series} {Experimental Methods in the Physical
  Sciences}, Vol.~\bibinfo {volume} {35}\ (\bibinfo  {publisher} {Academic
  Press, San Diego, USA},\ \bibinfo {year} {2013})\BibitemShut {NoStop}%
\bibitem [{\citenamefont {Lager}\ \emph {et~al.}(2008)\citenamefont {Lager},
  \citenamefont {Webb}, \citenamefont {Collins},\ and\ \citenamefont
  {Richmond}}]{Ian2008}%
  \BibitemOpen
  \bibfield  {author} {\bibinfo {author} {\bibfnamefont {Arnaud}\ \bibnamefont
  {Lager}}, \bibinfo {author} {\bibfnamefont {Kevin~John}\ \bibnamefont
  {Webb}}, \bibinfo {author} {\bibfnamefont {Ian~Ralph}\ \bibnamefont
  {Collins}}, \ and\ \bibinfo {author} {\bibfnamefont {Diane~Marie}\
  \bibnamefont {Richmond}},\ }\bibfield  {title} {\enquote {\bibinfo {title}
  {Losal enhanced oil recovery: Evidence of enhanced oil recovery at the
  reservoir scale},}\ }\href@noop {} {\bibfield  {journal} {\bibinfo  {journal}
  {SPE Journal}\ } (\bibinfo {year} {2008})}\BibitemShut {NoStop}%
\bibitem [{\citenamefont {Austad}\ \emph {et~al.}(2012)\citenamefont {Austad},
  \citenamefont {Shariatpanahi}, \citenamefont {Strand}, \citenamefont
  {Black},\ and\ \citenamefont {Webb}}]{Austad2012}%
  \BibitemOpen
  \bibfield  {author} {\bibinfo {author} {\bibfnamefont {T.}~\bibnamefont
  {Austad}}, \bibinfo {author} {\bibfnamefont {S.~F.}\ \bibnamefont
  {Shariatpanahi}}, \bibinfo {author} {\bibfnamefont {S.}~\bibnamefont
  {Strand}}, \bibinfo {author} {\bibfnamefont {C.~J.~J.}\ \bibnamefont
  {Black}}, \ and\ \bibinfo {author} {\bibfnamefont {K.~J.}\ \bibnamefont
  {Webb}},\ }\bibfield  {title} {\enquote {\bibinfo {title} {Conditions for a
  {Low}-{Salinity} {Enhanced} {Oil} {Recovery} ({EOR}) {Effect} in {Carbonate}
  {Oil} {Reservoirs}},}\ }\href {\doibase 10.1021/ef201435g} {\bibfield
  {journal} {\bibinfo  {journal} {Energy \& Fuels}\ }\textbf {\bibinfo {volume}
  {26}},\ \bibinfo {pages} {569--575} (\bibinfo {year} {2012})}\BibitemShut
  {NoStop}%
\bibitem [{\citenamefont {Hassenkam}\ \emph {et~al.}(2011)\citenamefont
  {Hassenkam}, \citenamefont {Pedersen}, \citenamefont {Dalby}, \citenamefont
  {Austad},\ and\ \citenamefont {Stipp}}]{Hassenkam2011}%
  \BibitemOpen
  \bibfield  {author} {\bibinfo {author} {\bibfnamefont {Tue}\ \bibnamefont
  {Hassenkam}}, \bibinfo {author} {\bibfnamefont {C.~S.}\ \bibnamefont
  {Pedersen}}, \bibinfo {author} {\bibfnamefont {K.}~\bibnamefont {Dalby}},
  \bibinfo {author} {\bibfnamefont {T.}~\bibnamefont {Austad}}, \ and\ \bibinfo
  {author} {\bibfnamefont {S.~L.~S.}\ \bibnamefont {Stipp}},\ }\bibfield
  {title} {\enquote {\bibinfo {title} {Pore scale observation of low salinity
  effects on outcrop and oil reservoir sandstone},}\ }\href@noop {} {\bibfield
  {journal} {\bibinfo  {journal} {Colloids and Surfaces A: Physicochemical and
  Engineering Aspects}\ }\textbf {\bibinfo {volume} {390}},\ \bibinfo {pages}
  {179--188} (\bibinfo {year} {2011})}\BibitemShut {NoStop}%
\bibitem [{\citenamefont {Orr}(2009)}]{Orr2009}%
  \BibitemOpen
  \bibfield  {author} {\bibinfo {author} {\bibfnamefont {F.~M.}\ \bibnamefont
  {Orr}},\ }\bibfield  {title} {\enquote {\bibinfo {title} {Onshore geologic
  storage of {$\textsf{CO}_2$}},}\ }\href@noop {} {\bibfield  {journal}
  {\bibinfo  {journal} {Science}\ }\textbf {\bibinfo {volume} {325}},\ \bibinfo
  {pages} {1656--1658} (\bibinfo {year} {2009})}\BibitemShut {NoStop}%
\bibitem [{\citenamefont {Bickle}(2009)}]{Bickle2009}%
  \BibitemOpen
  \bibfield  {author} {\bibinfo {author} {\bibfnamefont {Mike~J.}\ \bibnamefont
  {Bickle}},\ }\bibfield  {title} {\enquote {\bibinfo {title} {Geological
  carbon storage},}\ }\href@noop {} {\bibfield  {journal} {\bibinfo  {journal}
  {Nature Geoscience}\ }\textbf {\bibinfo {volume} {2}},\ \bibinfo {pages}
  {815--818} (\bibinfo {year} {2009})}\BibitemShut {NoStop}%
\bibitem [{\citenamefont {Singh}\ \emph {et~al.}(2017)\citenamefont {Singh},
  \citenamefont {Scholl}, \citenamefont {Brinkmann}, \citenamefont {DiMichiel},
  \citenamefont {Scheel}, \citenamefont {Herminghaus},\ and\ \citenamefont
  {Seemann}}]{Singh2017}%
  \BibitemOpen
  \bibfield  {author} {\bibinfo {author} {\bibfnamefont {Kamal}\ \bibnamefont
  {Singh}}, \bibinfo {author} {\bibfnamefont {Hagen}\ \bibnamefont {Scholl}},
  \bibinfo {author} {\bibfnamefont {Martin}\ \bibnamefont {Brinkmann}},
  \bibinfo {author} {\bibfnamefont {Marco}\ \bibnamefont {DiMichiel}}, \bibinfo
  {author} {\bibfnamefont {Mario}\ \bibnamefont {Scheel}}, \bibinfo {author}
  {\bibfnamefont {Stephan}\ \bibnamefont {Herminghaus}}, \ and\ \bibinfo
  {author} {\bibfnamefont {Ralf}\ \bibnamefont {Seemann}},\ }\bibfield  {title}
  {\enquote {\bibinfo {title} {The role of local instabilities in fluid
  invasion into permeable media},}\ }\href@noop {} {\bibfield  {journal}
  {\bibinfo  {journal} {Scientific Reports}\ }\textbf {\bibinfo {volume} {7}},\
  \bibinfo {pages} {444} (\bibinfo {year} {2017})}\BibitemShut {NoStop}%
\bibitem [{\citenamefont {Murison}\ \emph {et~al.}(2014)\citenamefont
  {Murison}, \citenamefont {Semin}, \citenamefont {Baret}, \citenamefont
  {Herminghaus}, \citenamefont {Schr\"oter},\ and\ \citenamefont
  {Brinkmann}}]{Murison2014}%
  \BibitemOpen
  \bibfield  {author} {\bibinfo {author} {\bibfnamefont {Julie}\ \bibnamefont
  {Murison}}, \bibinfo {author} {\bibfnamefont {Benoît}\ \bibnamefont
  {Semin}}, \bibinfo {author} {\bibfnamefont {Jean-Christophe}\ \bibnamefont
  {Baret}}, \bibinfo {author} {\bibfnamefont {Stephan}\ \bibnamefont
  {Herminghaus}}, \bibinfo {author} {\bibfnamefont {Matthias}\ \bibnamefont
  {Schr\"oter}}, \ and\ \bibinfo {author} {\bibfnamefont {Martin}\ \bibnamefont
  {Brinkmann}},\ }\bibfield  {title} {\enquote {\bibinfo {title} {Wetting
  heterogeneities in porous media control flow dissipation},}\ }\href@noop {}
  {\bibfield  {journal} {\bibinfo  {journal} {Phys. Rev. Applied}\ }\textbf
  {\bibinfo {volume} {2}},\ \bibinfo {pages} {034002} (\bibinfo {year}
  {2014})}\BibitemShut {NoStop}%
\bibitem [{\citenamefont {Blunt}\ and\ \citenamefont {King}(1991)}]{Blunt1991}%
  \BibitemOpen
  \bibfield  {author} {\bibinfo {author} {\bibfnamefont {Martin}\ \bibnamefont
  {Blunt}}\ and\ \bibinfo {author} {\bibfnamefont {Peter}\ \bibnamefont
  {King}},\ }\bibfield  {title} {\enquote {\bibinfo {title} {Relative
  permeabilities from two- and three-dimensional pore-scale network
  modelling},}\ }\href@noop {} {\bibfield  {journal} {\bibinfo  {journal}
  {Tran. Porous media Journal}\ }\textbf {\bibinfo {volume} {6}},\ \bibinfo
  {pages} {407--433} (\bibinfo {year} {1991})}\BibitemShut {NoStop}%
\bibitem [{\citenamefont {Blunt}(1997)}]{Blunt1997}%
  \BibitemOpen
  \bibfield  {author} {\bibinfo {author} {\bibfnamefont {Martin}\ \bibnamefont
  {Blunt}},\ }\bibfield  {title} {\enquote {\bibinfo {title} {Pore level
  modeling of the effects of wettability},}\ }\href@noop {} {\bibfield
  {journal} {\bibinfo  {journal} {SPE Journal}\ }\textbf {\bibinfo {volume}
  {2}},\ \bibinfo {pages} {494--510} (\bibinfo {year} {1997})}\BibitemShut
  {NoStop}%
\bibitem [{\citenamefont {Dixit}\ \emph {et~al.}(2000)\citenamefont {Dixit},
  \citenamefont {Buckley}, \citenamefont {McDougall},\ and\ \citenamefont
  {Sorbie}}]{Dixit2000}%
  \BibitemOpen
  \bibfield  {author} {\bibinfo {author} {\bibfnamefont {A.~B.}\ \bibnamefont
  {Dixit}}, \bibinfo {author} {\bibfnamefont {J.~S.}\ \bibnamefont {Buckley}},
  \bibinfo {author} {\bibfnamefont {S.~R.}\ \bibnamefont {McDougall}}, \ and\
  \bibinfo {author} {\bibfnamefont {K.~S.}\ \bibnamefont {Sorbie}},\ }\bibfield
   {title} {\enquote {\bibinfo {title} {Empirical measures of wettability in
  porous media and the relationship between them derived from pore-scale
  modelling},}\ }\href@noop {} {\bibfield  {journal} {\bibinfo  {journal}
  {Tran. Porous media Journal}\ }\textbf {\bibinfo {volume} {40}},\ \bibinfo
  {pages} {27--54} (\bibinfo {year} {2000})}\BibitemShut {NoStop}%
\bibitem [{\citenamefont {{\O}ren}\ and\ \citenamefont
  {Bakke}(2003)}]{Oren2003}%
  \BibitemOpen
  \bibfield  {author} {\bibinfo {author} {\bibfnamefont {P{\aa}l-Eric}\
  \bibnamefont {{\O}ren}}\ and\ \bibinfo {author} {\bibfnamefont {Stig}\
  \bibnamefont {Bakke}},\ }\bibfield  {title} {\enquote {\bibinfo {title}
  {Reconstruction of {Berea} sandstone and pore-scale modelling of wettability
  effects},}\ }\href@noop {} {\bibfield  {journal} {\bibinfo  {journal} {SPE
  Journal}\ }\textbf {\bibinfo {volume} {39}},\ \bibinfo {pages} {177--199}
  (\bibinfo {year} {2003})}\BibitemShut {NoStop}%
\bibitem [{\citenamefont {Zhao}\ \emph {et~al.}(2010)\citenamefont {Zhao},
  \citenamefont {Blunt},\ and\ \citenamefont {Yao}}]{Zhao2010}%
  \BibitemOpen
  \bibfield  {author} {\bibinfo {author} {\bibfnamefont {Xiucai}\ \bibnamefont
  {Zhao}}, \bibinfo {author} {\bibfnamefont {Martin~J.}\ \bibnamefont {Blunt}},
  \ and\ \bibinfo {author} {\bibfnamefont {Jun}\ \bibnamefont {Yao}},\
  }\bibfield  {title} {\enquote {\bibinfo {title} {Pore-scale modeling: Effects
  of wettability on waterflood oil recovery},}\ }\href@noop {} {\bibfield
  {journal} {\bibinfo  {journal} {Journal of Petroleum Science and
  Engineering}\ }\textbf {\bibinfo {volume} {71}},\ \bibinfo {pages} {169--178}
  (\bibinfo {year} {2010})}\BibitemShut {NoStop}%
\bibitem [{\citenamefont {Fatt}\ \emph {et~al.}(1959)\citenamefont {Fatt},
  \citenamefont {Waldemar},\ and\ \citenamefont {Klikoff}}]{Fatt1959}%
  \BibitemOpen
  \bibfield  {author} {\bibinfo {author} {\bibfnamefont {I.}~\bibnamefont
  {Fatt}}, \bibinfo {author} {\bibfnamefont {A.}~\bibnamefont {Waldemar}}, \
  and\ \bibinfo {author} {\bibfnamefont {Jr.}\ \bibnamefont {Klikoff}},\
  }\bibfield  {title} {\enquote {\bibinfo {title} {Effect of fractional
  wettability on multiphase flow through porous media},}\ }\href@noop {}
  {\bibfield  {journal} {\bibinfo  {journal} {SPE Journal}\ }\textbf {\bibinfo
  {volume} {11}},\ \bibinfo {pages} {71--75} (\bibinfo {year}
  {1959})}\BibitemShut {NoStop}%
\bibitem [{\citenamefont {Lessel}\ \emph {et~al.}(2014)\citenamefont {Lessel},
  \citenamefont {B\"aumchen}, \citenamefont {Klos}, \citenamefont {H\"ahl},
  \citenamefont {Fetzer}, \citenamefont {Paulus}, \citenamefont {Seemann},\
  and\ \citenamefont {Jacobs}}]{Lessel2014}%
  \BibitemOpen
  \bibfield  {author} {\bibinfo {author} {\bibfnamefont {M.}~\bibnamefont
  {Lessel}}, \bibinfo {author} {\bibfnamefont {O.}~\bibnamefont {B\"aumchen}},
  \bibinfo {author} {\bibfnamefont {M.}~\bibnamefont {Klos}}, \bibinfo {author}
  {\bibfnamefont {H.}~\bibnamefont {H\"ahl}}, \bibinfo {author} {\bibfnamefont
  {R.}~\bibnamefont {Fetzer}}, \bibinfo {author} {\bibfnamefont
  {M.}~\bibnamefont {Paulus}}, \bibinfo {author} {\bibfnamefont
  {R.}~\bibnamefont {Seemann}}, \ and\ \bibinfo {author} {\bibfnamefont
  {K.}~\bibnamefont {Jacobs}},\ }\bibfield  {title} {\enquote {\bibinfo {title}
  {Self-assembled silane monolayers: an efficient step-by-step recipe for
  high-quality, low energy surfaces},}\ }\href@noop {} {\bibfield  {journal}
  {\bibinfo  {journal} {Surface and Interface Analysis}\ }\textbf {\bibinfo
  {volume} {47}},\ \bibinfo {pages} {557–564} (\bibinfo {year}
  {2014})}\BibitemShut {NoStop}%
\bibitem [{\citenamefont {Perrin}\ \emph {et~al.}(2016)\citenamefont {Perrin},
  \citenamefont {Lhermerout}, \citenamefont {Davitt}, \citenamefont {Rolley},\
  and\ \citenamefont {Andreotti}}]{Perrin2016}%
  \BibitemOpen
  \bibfield  {author} {\bibinfo {author} {\bibfnamefont {H.}~\bibnamefont
  {Perrin}}, \bibinfo {author} {\bibfnamefont {R.}~\bibnamefont {Lhermerout}},
  \bibinfo {author} {\bibfnamefont {K.}~\bibnamefont {Davitt}}, \bibinfo
  {author} {\bibfnamefont {E.}~\bibnamefont {Rolley}}, \ and\ \bibinfo {author}
  {\bibfnamefont {B.}~\bibnamefont {Andreotti}},\ }\bibfield  {title} {\enquote
  {\bibinfo {title} {Defects at the nanoscale impact contact line motion at all
  scales},}\ }\href@noop {} {\bibfield  {journal} {\bibinfo  {journal} {Phys.
  Rev. Lett.}\ }\textbf {\bibinfo {volume} {116}},\ \bibinfo {pages} {184502}
  (\bibinfo {year} {2016})}\BibitemShut {NoStop}%
\bibitem [{\citenamefont {Dufour}\ \emph {et~al.}(2016)\citenamefont {Dufour},
  \citenamefont {Semprebon},\ and\ \citenamefont {Herminghaus}}]{Dufour2016}%
  \BibitemOpen
  \bibfield  {author} {\bibinfo {author} {\bibfnamefont {R.}~\bibnamefont
  {Dufour}}, \bibinfo {author} {\bibfnamefont {C.}~\bibnamefont {Semprebon}}, \
  and\ \bibinfo {author} {\bibfnamefont {S.}~\bibnamefont {Herminghaus}},\
  }\bibfield  {title} {\enquote {\bibinfo {title} {Filling transitions on rough
  surfaces: inadequacy of {Gaussian} surface models},}\ }\href@noop {}
  {\bibfield  {journal} {\bibinfo  {journal} {Phys. Rev. E}\ }\textbf {\bibinfo
  {volume} {93}},\ \bibinfo {pages} {032802} (\bibinfo {year}
  {2016})}\BibitemShut {NoStop}%
\bibitem [{\citenamefont {Wylock}\ \emph {et~al.}(2012)\citenamefont {Wylock},
  \citenamefont {Pradas}, \citenamefont {Haut}, \citenamefont {Colinet},\ and\
  \citenamefont {Kalliadasis}}]{Wylock2012}%
  \BibitemOpen
  \bibfield  {author} {\bibinfo {author} {\bibfnamefont {C.}~\bibnamefont
  {Wylock}}, \bibinfo {author} {\bibfnamefont {M.}~\bibnamefont {Pradas}},
  \bibinfo {author} {\bibfnamefont {B.}~\bibnamefont {Haut}}, \bibinfo {author}
  {\bibfnamefont {P.}~\bibnamefont {Colinet}}, \ and\ \bibinfo {author}
  {\bibfnamefont {S.}~\bibnamefont {Kalliadasis}},\ }\bibfield  {title}
  {\enquote {\bibinfo {title} {Disorder-induced hysteresis and nonlocality of
  contact line motion in chemically heterogeneous microchannels},}\ }\href@noop
  {} {\bibfield  {journal} {\bibinfo  {journal} {Physics of Fluids}\ }\textbf
  {\bibinfo {volume} {24}},\ \bibinfo {pages} {032108} (\bibinfo {year}
  {2012})}\BibitemShut {NoStop}%
\bibitem [{\citenamefont {Kohonen}\ \emph {et~al.}(2004)\citenamefont
  {Kohonen}, \citenamefont {Geromichalos}, \citenamefont {Scheel},
  \citenamefont {Schier},\ and\ \citenamefont {Herminghaus}}]{Kohonen2004}%
  \BibitemOpen
  \bibfield  {author} {\bibinfo {author} {\bibfnamefont {Mika~M.}\ \bibnamefont
  {Kohonen}}, \bibinfo {author} {\bibfnamefont {Dimitrios}\ \bibnamefont
  {Geromichalos}}, \bibinfo {author} {\bibfnamefont {Mario}\ \bibnamefont
  {Scheel}}, \bibinfo {author} {\bibfnamefont {Christoph}\ \bibnamefont
  {Schier}}, \ and\ \bibinfo {author} {\bibfnamefont {Stephan}\ \bibnamefont
  {Herminghaus}},\ }\bibfield  {title} {\enquote {\bibinfo {title} {On
  capillary bridges in wet granular materials},}\ }\href@noop {} {\bibfield
  {journal} {\bibinfo  {journal} {Physica A}\ }\textbf {\bibinfo {volume}
  {339}},\ \bibinfo {pages} {7--15} (\bibinfo {year} {2004})}\BibitemShut
  {NoStop}%
\bibitem [{\citenamefont {Powell}(1979)}]{Powell1979}%
  \BibitemOpen
  \bibfield  {author} {\bibinfo {author} {\bibfnamefont {M.~J.}\ \bibnamefont
  {Powell}},\ }\bibfield  {title} {\enquote {\bibinfo {title} {Site percolation
  in randomly packed spheres},}\ }\href@noop {} {\bibfield  {journal} {\bibinfo
   {journal} {Phys. Rev. B}\ }\textbf {\bibinfo {volume} {20}},\ \bibinfo
  {pages} {4194} (\bibinfo {year} {1979})}\BibitemShut {NoStop}%
\bibitem [{\citenamefont {Scheel}\ \emph
  {et~al.}(2008{\natexlab{a}})\citenamefont {Scheel}, \citenamefont {Seemann},
  \citenamefont {Brinkmann}, \citenamefont {DiMichiel}, \citenamefont
  {Sheppard}, \citenamefont {Breidenbach},\ and\ \citenamefont
  {Herminghaus}}]{Scheel2008}%
  \BibitemOpen
  \bibfield  {author} {\bibinfo {author} {\bibfnamefont {Mario}\ \bibnamefont
  {Scheel}}, \bibinfo {author} {\bibfnamefont {Ralf}\ \bibnamefont {Seemann}},
  \bibinfo {author} {\bibfnamefont {Martin}\ \bibnamefont {Brinkmann}},
  \bibinfo {author} {\bibfnamefont {Marco}\ \bibnamefont {DiMichiel}}, \bibinfo
  {author} {\bibfnamefont {Adrian}\ \bibnamefont {Sheppard}}, \bibinfo {author}
  {\bibfnamefont {Boris}\ \bibnamefont {Breidenbach}}, \ and\ \bibinfo {author}
  {\bibfnamefont {Stephan}\ \bibnamefont {Herminghaus}},\ }\bibfield  {title}
  {\enquote {\bibinfo {title} {Morphological clues to wet granular pile
  stability},}\ }\href@noop {} {\bibfield  {journal} {\bibinfo  {journal}
  {Nature Materials}\ }\textbf {\bibinfo {volume} {7}},\ \bibinfo {pages} {189}
  (\bibinfo {year} {2008}{\natexlab{a}})}\BibitemShut {NoStop}%
\bibitem [{\citenamefont {Scheel}\ \emph
  {et~al.}(2008{\natexlab{b}})\citenamefont {Scheel}, \citenamefont {Seemann},
  \citenamefont {Brinkmann}, \citenamefont {DiMichiel}, \citenamefont
  {Sheppard},\ and\ \citenamefont {Herminghaus}}]{Scheel2008a}%
  \BibitemOpen
  \bibfield  {author} {\bibinfo {author} {\bibfnamefont {Mario}\ \bibnamefont
  {Scheel}}, \bibinfo {author} {\bibfnamefont {Ralf}\ \bibnamefont {Seemann}},
  \bibinfo {author} {\bibfnamefont {Martin}\ \bibnamefont {Brinkmann}},
  \bibinfo {author} {\bibfnamefont {Marco}\ \bibnamefont {DiMichiel}}, \bibinfo
  {author} {\bibfnamefont {Adrian}\ \bibnamefont {Sheppard}}, \ and\ \bibinfo
  {author} {\bibfnamefont {Stephan}\ \bibnamefont {Herminghaus}},\ }\bibfield
  {title} {\enquote {\bibinfo {title} {Liquid distribution and cohesion in wet
  granular assemblies beyond the capillary bridge regime},}\ }\href@noop {}
  {\bibfield  {journal} {\bibinfo  {journal} {J. Phys.: Cond. Mat.}\ }\textbf
  {\bibinfo {volume} {20}},\ \bibinfo {pages} {494236} (\bibinfo {year}
  {2008}{\natexlab{b}})}\BibitemShut {NoStop}%
\bibitem [{\citenamefont {Herminghaus}(2013)}]{Herminghaus2013}%
  \BibitemOpen
  \bibfield  {author} {\bibinfo {author} {\bibfnamefont {Stephan}\ \bibnamefont
  {Herminghaus}},\ }\href@noop {} {\emph {\bibinfo {title} {Wet Granular
  Matter: a Truly Complex Fluid}}},\ \bibinfo {series} {Series in Soft
  Condensed Matter}, Vol.~\bibinfo {volume} {6}\ (\bibinfo  {publisher} {World
  Scientific},\ \bibinfo {year} {2013})\BibitemShut {NoStop}%
\end{thebibliography}

\end{document}